# Recovering pyramid WS gain in non-common path aberration correction mode via deformable lens

D. Magrin[a,b], S. Bonora[c], M. Quintavalla[c], P. Favazza[d], M. Bergomi[a,b],
G. Umbriaco[a,b,d], S. Chinellato[a,b], R. Ragazzoni[a,b].

[a]INAF - Osservatorio Astronomico di Padova, Vicolo dell'Osservatorio 5, 35122 Padova, Italy
[b]ADONI –Laboratorio Nazionale di Ottica Adattiva
[c]CNR-IFN – Istituto di Fotonica e Nanotecnologie, via Trasea 7, 35131 Padova, Italy
[d]Dipartimento di Fisica ed Astronomia -Università degli Studi di Padova, Vicolo dell'Osservatorio 3, 35122 Padova, Italy

**ABSTRACT**

It is by now well known that pyramid based wavefront sensors, once in closed loop, have the capability to improve more and more the gain as the reference natural star image size is getting smaller on the pyramid pin. Especially in extreme adaptive optics applications, in order to correct the non-common path aberrations between the scientific and sensing channel, it is common use to inject a certain amount of offset wavefront deformation into the DM(s), departing at the same time the pyramid from the optimal working condition. In this paper we elaborate on the possibility to correct the low order non-common path aberrations at the pyramid wavefront sensor level by means of an adaptive refractive lens placed on the optical path before the pyramid itself, allowing the mitigation of the gain loss.

## 1. INTRODUCTION

It is about 20 years since Pyramid Wavefront Sensor (PWS) was firstly proposed[1] and then demonstrated its potentiality on sky [2, 3] at the Telescopio Nazionale Galileo in the Canary Islands. Nowadays, PWS are utilized in several observatories [4, 5, 6] all over the world and have been selected as baseline in many incoming instrumentations [7].

Despite some technical advantages, such as the possibility to bin the detector pixels as a function of corrected modes, PWS superiority with respect to previously more utilized Shack-Hartmann wavefront sensors is probably due to its capability to improve its sensitivity as the image of the reference natural star gets smaller, i.e. as the closed loop iterations increase [8,9]. Such virtuous loop could potentially lead to outstanding results as demonstrated recently by the First Light Adaptive Optics system at Large Binocular Telescope [10].

Never the less, the performance improvement of the PWS could be in principle reduced, or in the worst cases even nullify, by the so called non-common path aberrations (NCPA). Basically, NCPA appears when the instrument and the wavefront sensing arm and the instrument experience different aberrations between them, which could originate from design mismatching, poor relative alignment but also from gravity and thermal differential effects.

The usual way to correct for such NCPA is to introduce an offset in the wavefront sensor signals that corresponds to the aberration to correct. In such a way, when the AO loop is closed, the Deformable Mirror (DM) will converge to the shape required to null the NCPA on the instrument. Having PWS a restricted linear range, this correction requires even a more sophisticated approach [11]. In any case, the effect to correct the NCPA by the DM, while it increases the quality on the instrument, it decreases the quality of the star imaged on the pin of the pyramid moving it intrinsically away from the best regime. Depending on the severity of the NCPA magnitude, this effect could lead to performance loss when in the closed loop in terms of delivered Strehl ratio [12].

We propose here to insert a deformable refractive element, a deformable lens, on the optical path of the PWS before the pyramid which function is aimed to correct the low order NCPA at the wavefront sensor level allowing the mitigation of performance loss.

## 2. THE REFRACTIVE DEFORMABLE ELEMENT

The Adaptive Lenses (AL) have already demonstrated their potentiality in different fields, such as microscopy in vivo ophthalmic imaging applications [13].

The AL prototype (see Figure 1), that we manufactured and tested, is made up by two N-BK7 optical windows (150 μm thick) filled with transparent liquid, matching the refractive index of the glass. On the outer perimeter of each window piezoelectric (PZT) ring actuators are glued. Both rings are divided into eight sectors that can be actuated independently. The piezoelectric actuators have an external diameter of 25 mm and an internal one of 10 mm with a thickness of 200 μm. The area inside the ring actuator is the clear aperture of the deformable lens and has a diameter of 10 mm. Because the PZT is rigidly glued to the glass disc, its expansion or contraction means that, in first approximation, the unimorph couple formed by the glass and the PZT deforms the glass window as a paraboloid. The local surface lens shape deformation is proportional to the applied voltage.

One of the two windows is mounted on a rigid aluminum ring which is necessary to change the boundary condition and to move the maximum (minimum) of the window deformation inside the clear aperture. Therefore the activation of electrodes in the two windows gives a different lens shape. In addition, to help the lens in the generation of spherical aberration the central part of the lens is stiffened with a glass disc with the refractive index matched to the one of the liquid. The current transmissivity on the visible band is about 94%. It has to be underlined that at the moment the windows external surfaces are not AR coated, probably accounting for most of the throughput losses involved in this prototype.

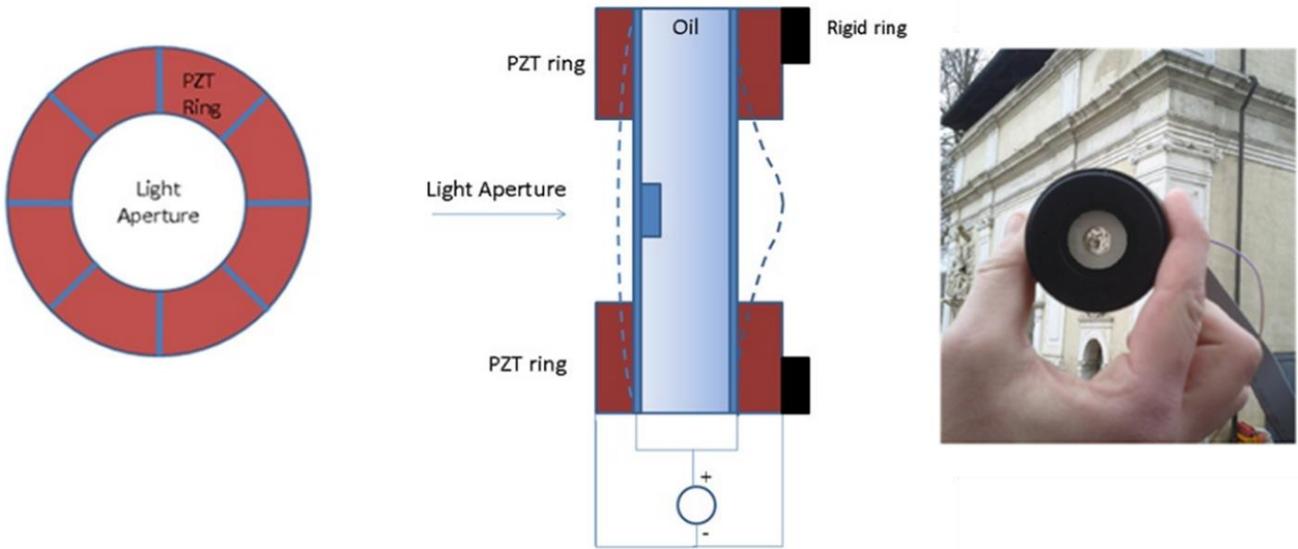

Figure 1: Scheme of the Adaptive Lens. Left side: actuator arrangement. Center: cross view of the AL. Right side: picture of the AL.

The wavefront of a collimated beam passing through the AL is basically given by the difference of the shape of the two optical windows multiplied by the refractive index of the transparent liquid contained on the lens. The shape of each window is the determined by the applied voltages and can be well approximated by the algebraic sum of the deformation induced by each single electrode in both windows. The actuators are controlled using a 18 channel high voltage (+/-125 V) driver. The maximum allowed frequency is about 200 Hz. The deformation principle is shown in Figure 2.

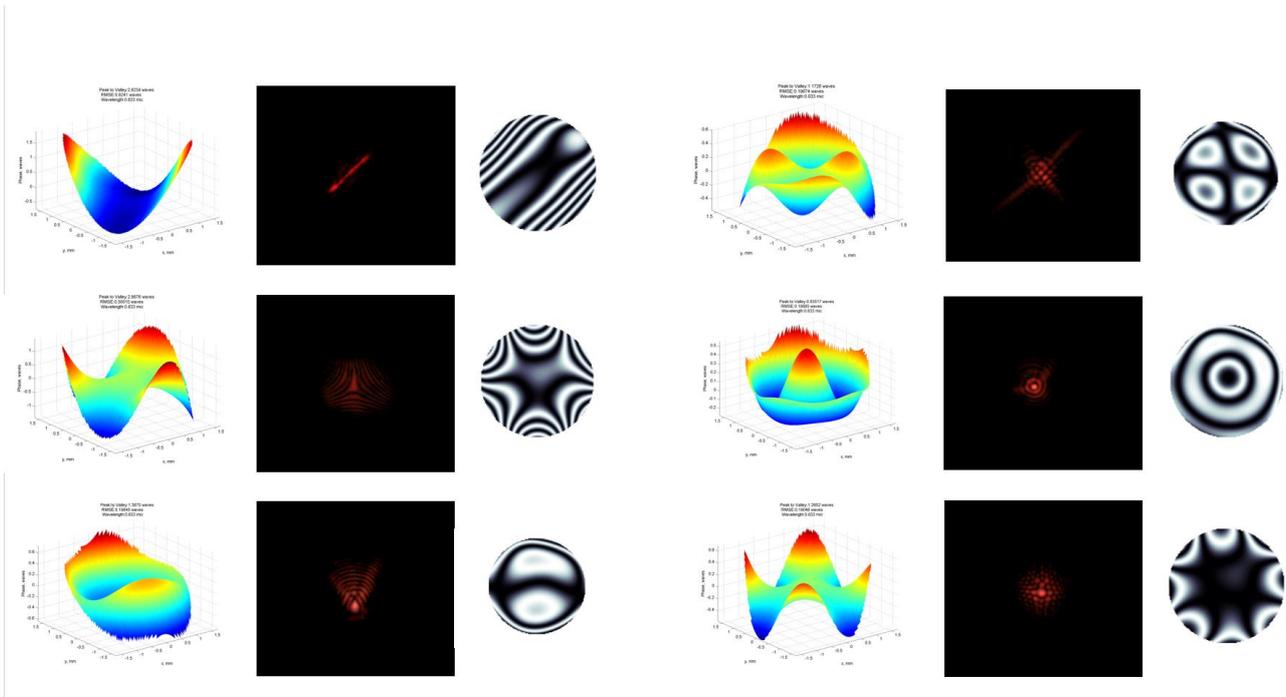

Figure 2 Scheme of the AL deformations principle.

## 3. PRELIMINARY PROTOTYPE TESTING

A preliminary test and calibration campaign has been conducted on three AL prototypes. These firstly manufactured prototypes have not still been tuned for the proposed astronomical application, but they have been used to increase the image quality delivered in microscope images. In particular, the dynamical range presented by such AL is quite large reducing the sensitivity to small applied deformations.

In the calibration set-up, a collimated beam at $\mu=633$ nm passed through the AL and the effects of the AL deformations were measured by mean of a Shack-Hartmann wavefront sensor. The AL was then characterized in order to be able to generate the first fifteen Zernike polynomials and their combination. To generate the polynomial we started from the measurements of the influence functions and then we retrieved the control voltages necessary for their generation from the pseudoinverse matrix.

We have tested three different prototypes: an AL with clear aperture 10 mm (L10A), an AL with clear aperture 22 mm (L22A) and an AL with clear aperture 10 mm with a different transparent liquid (L10B). The wavefront amplitude and the residual errors in open loop for each case are shown in Figure 3.

We measured a maximum stroke of about 10 waves for the cases L10A and L10B and of about 18 waves for L22A. The raise time (10%-90%) for the L10A and L10B cases resulted to be 1.5 ms while it is 3 ms for the L22A.

In open loop, the residual errors are always below 0.2 waves for all the Zernike polynomials. We measured the residual errors also in closed loop obtaining values below 0.08 waves. Such vales are still not sufficiently suitable for the application we proposed here. However, given the fact that the experimental set-up was not properly characterized, such numbers have to be intended as upper limits. Moreover, it is our intention to manufacture an ad-hoc prototype with limited dynamical range (1 wave) and increased shape deformation precision.

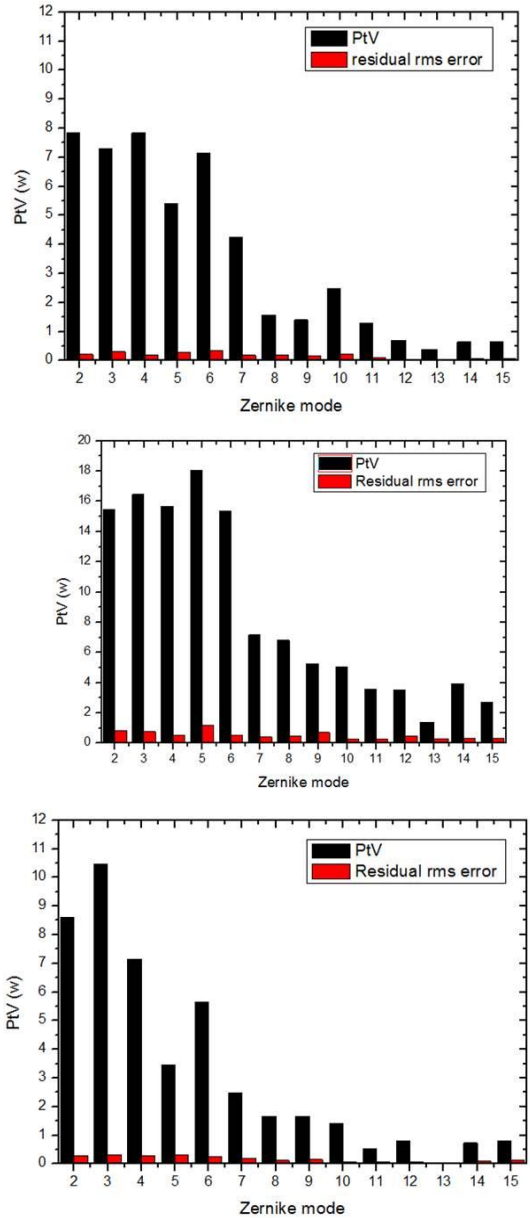

Figure 3 Wavefront amplitude (black) and the residual errors (red) for L10A (top), L22A (middle), L10B (bottom).

## 4. NCPA CORRECTION WITH ADAPTIVE LENS

The NCPA correction is typically achieved in a closed loop system by introducing on the deformable mirror the wavefront differences between the instrument and the wavefront sensor. In the case of PWS, the drawback is to force the pyramid out of its optimal working regime, and, depending on the severity of the NCPA amplitude, reducing the overall performance of the AO system.

The key idea, we propose here, is to correct the offset low-order wavefront deformation seen by the pyramid wavefront sensor, through an adaptive lens, placed in front of the pyramid itself, which will introduce the opposite deformation on the wavefront (see Figure 4). Usually the NCPA are measured when the system is in closed loop and the PWS are in the best regime, by estimating the image quality on the scientific detector of the instrument. Then the DM is piloted in order to increase the performances on the latter detector. In our scheme, the opposite deformation should be given to the AL so that in principle allows the full recovery of the pyramid gain even in NCPA correction mode. In this case the AL is piloted in open loop. For non-static NCPA, the correction can take advantages of pre-calibrated lookup tables in different observing conditions.

In case of instrumentation that includes an internal wavefront sensing system, such as SHARK-NIR [14], able to continuously monitor in real time the image quality, it is in principle possible to correct the NCPA on the DM and counter correct the wavefront on the PWS by AL in closed loop. Typically, NCPA time evolution is small and the correction does not require a high frequency rate. The current AL 200Hz frequency is suitable for such application. However, if the frequency rate could be extended to 500-1000 Hz in the next prototype version, such device could be used also for pyramid modulation.

The refractive nature, compactness and the clear aperture sizes of the AL make this device suitable for an easy installation on board of already existing PWS systems without the need of complex re-design or insertion of folding mirrors.

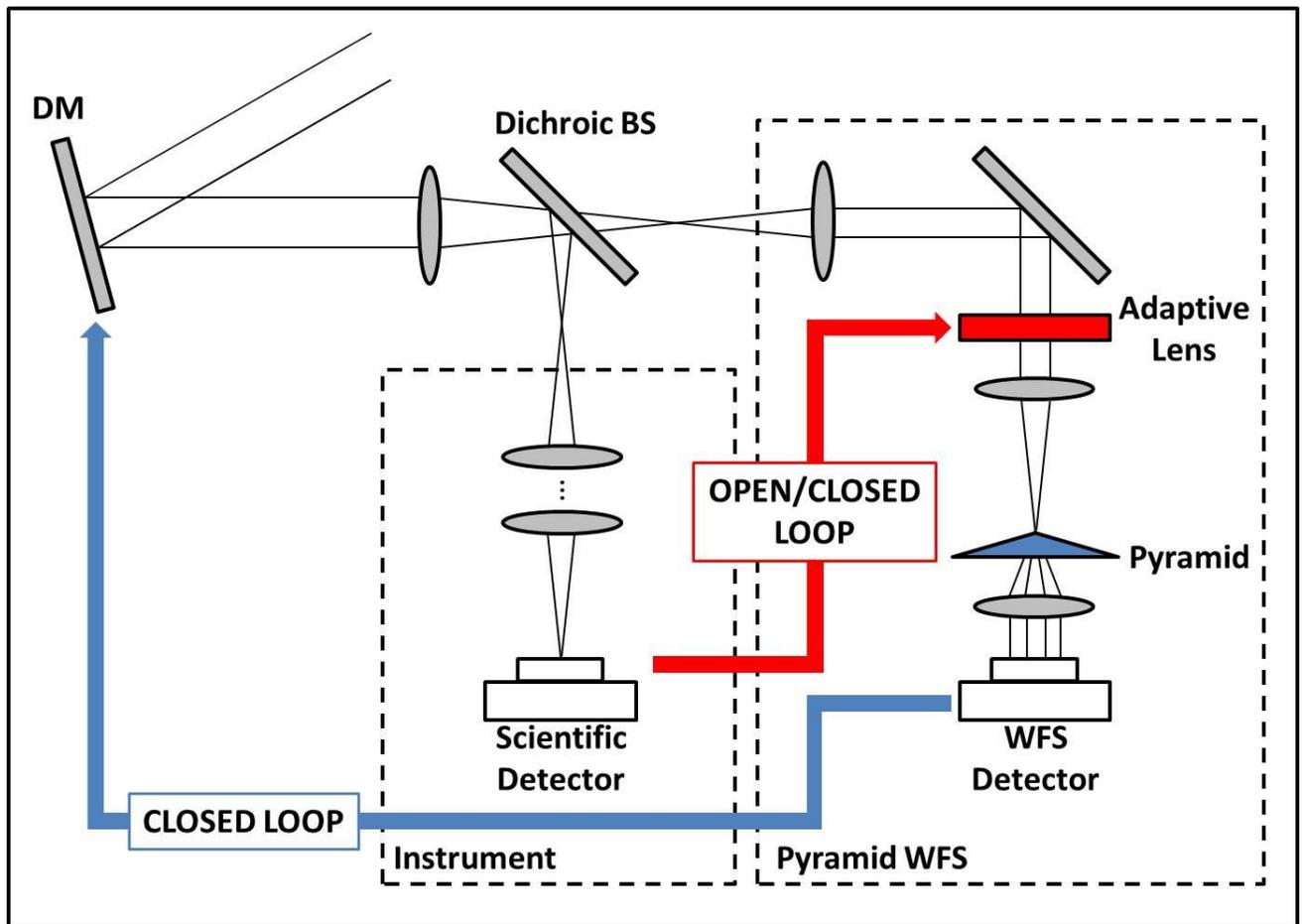

Figure 4 Scheme of the NCPA correction with AL placed in front of the pyramid.

## 5. FUTURE STEPS

In order to demonstrate the concept and its feasibility, we have individuated the following path:

- In Lab demonstration of the concept. Following the scheme in Figure 4, we are working on a setup where the instrument channel is substituted by a Shak-Hartman wavefront sensor. In a first phase, the DM is not present and we will try to minimize the NCPA between the two channels in open loop. In a second phase, we will introduce the DM in order to test the system in real-time both in open and closed loop.
- Manufacturing of an ad hoc lens. From the already known results and the ones retrieved from the previous step we will manufacture a new AL fine-tuned for this application. The new prototype will be fully characterized in terms of transmissivity, chromatic aberrations, stroke and correction residuals.
- On sky demonstration of the concept. At the time being we foresee to test the system on sky exploiting the facility at the Coudé focus of the Telescopio Copernico in Asiago [15], that is actually under installation and will be dedicated to such kind of activities, in the framework of ADONI.

After the on sky concept demonstration, we will propose to install the AL on an already existing wavefront sensor system based on pyramid.